\documentclass[preprintnumbers,amsmath,amssymbm,preprint]{revtex4}
\usepackage{epsfig}
\usepackage{graphicx}
\usepackage{amssymb}

\begin{document}

\title{Curved spacetimes with continuous light disks}
\author{Shahar Hod}
\address{The Ruppin Academic Center, Emeq Hefer 40250, Israel}
\address{}
\address{The Jerusalem Multidisciplinary Institute, Jerusalem 91010, Israel}
\date{\today}

\begin{abstract}
\ \ \ Highly curved spacetimes of compact astrophysical objects are known to possess light 
rings (null circular geodesics) with {\it discrete} radii 
on which massless particles can perform closed circular motions. 
In the present compact paper, we reveal for the first time 
the existence of isotropic curved spacetimes that possess light disks 
which are made of a {\it continuum} of closed light rings. 
In particular, using analytical techniques which are based on the non-linearly coupled 
Einstein-matter field equations, we prove that 
these physically intriguing spacetimes contain a central compact core of radius $r_->0$ that 
supports an outer spherical shell with an infinite number (a continuum) 
of null circular geodesic which are all 
characterized by the functional relations $4\pi r^2_{\gamma}p(r_{\gamma})=1-3m(r_{\gamma})/r_{\gamma}$ 
and $8\pi r^2_{\gamma}(\rho+p)=1$ for $r_{\gamma}\in[r_-,r_+]$ [here 
$\{\rho,p\}$ are respectively 
the energy density and the isotropic pressure of the self-gravitating matter fields and 
$m(r)$ is the gravitational mass contained within the sphere of radius $r$].  
\end{abstract}
\bigskip
\maketitle

\section{Introduction}
Observational studies \cite{Aki} have recently confirmed that, 
in accord with the predictions of general relativity \cite{Bar,Chan,Shap,Hodub,Herne,Hodrec,YY}, 
closed null circular geodesics exist in highly curved spacetimes of self-gravitating compact objects. 
Interestingly, it is well established in the physics literature (see 
\cite{Bar,Chan,Shap,Herne,Hodns,Hodrec,YY,Aki,Mash,Goeb,Hod1,Dec,Hodhair,Hodfast,YP,
Hodm,Hodub,Lu1,Hodlwd,Pod,Ame,Ste} and 
references therein) that the presence of light rings, 
on which massless particles can perform closed circular motions in curved spacetimes, 
has many important implications on the physical, observational, and mathematical properties of the 
corresponding central black holes and horizonless compact objects. 

For example, the nearly circular (slightly perturbed) motions of massless fields along unstable null geodesics of curved 
spacetimes determine, in the eikonal (short wavelength) regime, the characteristic relaxation timescales of the correspoinding perturbed central compact objects \cite{Mash,Goeb,Hod1,Dec}. 
In addition, it has been proved \cite{Hodfast,YP} that, as measured by asymptotic observers, 
the equatorial light ring of a curved black-hole spacetime determines the shortest possible orbital period around the 
central black hole. 

Moreover, the presence of light rings around central compact objects is known to determine 
their optical properties as measured by far away observers \cite{Pod,Ame,Ste}. 
In addition, it has been revealed \cite{Hodhair,Hodub,Hodlwd,Hod1} that the radius of the 
innermost light ring (the radius of the innermost null circular geodesic) of a hairy curved 
black-hole spacetime provides a physically interesting lower bound on the effective radial lengths 
of the supported self-gravitating hairy matter configurations. 
   
Motivated by the fact that light rings (null circular geodesics) are an important ingredient of 
highly curved spacetimes that describe black holes and horizonless compact objects \cite{Bar,Chan,Shap,Hodub,Herne,Hodrec,YY,Aki}, 
in the present compact paper we raise the following physically interesting question: 
Is it possible to build curved spacetimes that contain an {\it infinite} number of light rings?

Using the non-linearly coupled Einstein-matter field equations, in the present paper we shall reveal 
the physically intriguing fact that the answer to the above stated question is `yes'! 
In particular, we shall explicitly prove below that there are non-trivial curved spacetimes that possess 
spherical shells $r_{\gamma}\in[r_-,r_+]$ with $r_->0$ which contain an infinite number (a {\it continuum}) of light rings.

\section{Description of the system}
We shall study, using analytical techniques, the physical and mathematical properties of self-gravitating isotropic 
matter configurations that possess closed light rings. 
Using the Schwarzschild-like spacetime coordinates $\{t,r,\theta,\phi\}$ one can express 
the line element of the corresponding spherically symmetric curved spacetimes in the form 
\cite{Hodfast,Hodm,Noteunit}
\begin{equation}\label{Eq1}
ds^2=-e^{-2\delta}\mu dt^2 +\mu^{-1}dr^2+r^2(d\theta^2 +\sin^2\theta d\phi^2)\  ,
\end{equation}
where $\mu(r)$ and $\delta(r)$ are radially dependent dimensionless metric functions. 

The non-linearly coupled Einstein-matter field 
equations $G^{\mu}_{\nu}=8\pi T^{\mu}_{\nu}$ can be expressed in the form \cite{Hodfast,Hodm} 
\begin{equation}\label{Eq2}
{{d\mu}\over{dr}}=-8\pi r\rho+{{1-\mu}\over{r}}\
\end{equation}
and
\begin{equation}\label{Eq3}
{{d\delta}\over{dr}}=-{{4\pi r(\rho +p)}\over{\mu}}\  ,
\end{equation}
where \cite{Bond1}
\begin{equation}\label{Eq4}
\rho\equiv-T^{t}_{t}\ \ \ \ \text{and}\ \ \ \ p\equiv T^{r}_{r}=T^{\theta}_{\theta}=T^{\phi}_{\phi}\
\end{equation}
are respectively the energy density and the isotropic pressure of 
the self-gravitating matter fields. 

A curved spacetime with a regular origin is characterized by the relations \cite{Bekreg}
\begin{equation}\label{Eq5}
\mu(r=0)=1+O(r^2)\
\end{equation}
and
\begin{equation}\label{Eq6}
\delta(r=0)<\infty\  .
\end{equation}
In addition, an asymptotically flat spacetime is characterized by the relations \cite{Bekreg,NoteMM}
\begin{equation}\label{Eq7}
\mu(r\to\infty)\to1+O(M/r)\ 
\end{equation}
and
\begin{equation}\label{Eq8}
\delta(r\to\infty)\to0\  .
\end{equation} 

The dimensionless metric function $\mu(r)$ can be expressed, 
using the Einstein differential equation (\ref{Eq2}), in the compact mathematical form 
\begin{equation}\label{Eq9}
\mu(r)=1-{{2m(r)}\over{r}}\  ,
\end{equation}
where 
\begin{equation}\label{Eq10}
m(r)=\int_{0}^{r} 4\pi x^{2}\rho(x)dx\
\end{equation}
is the gravitational mass contained within a sphere of radius $r$. 

Taking cognizance of Eqs. (\ref{Eq5}), (\ref{Eq9}), and (\ref{Eq10}) one deduces that the density function of the 
self-gravitating matter fields is characterized by the near-origin functional relation 
\begin{equation}\label{Eq11}
\rho(r)<\infty\ \ \ \ \text{for}\ \ \ \ r\to0\  .
\end{equation}
In addition, taking cognizance of Eqs. (\ref{Eq7}), (\ref{Eq9}), and (\ref{Eq10}) one finds that the density function is 
characterized by the asymptotic radial behavior 
\begin{equation}\label{Eq12}
r^3\rho(r)\to0\ \ \ \ \text{for}\ \ \ \ r\to\infty\  .
\end{equation}

\section{Isotropic curved spacetimes that possess light disks}
In the present section we shall use the non-linearly coupled Einstein-matter field equations in order to reveal 
the intriguing existence of 
curved spacetimes that contain central cores of radius $r_->0$ 
which support matter shells $r\in[r_-,r_+]$ that contain {\it infinite} sequences 
of null circular geodesics (continua of closed light rings).

\subsection{The radial locations of light rings in curved spacetimes}
The radial locations of null circular geodesics in the spherically symmetric curved spacetime (\ref{Eq1}) 
are determined by the dimensionless functional relation \cite{Hodhair}
\begin{equation}\label{Eq13}
{\cal N}(r)\equiv 3\mu-1-8\pi r^2p=0\ \ \ \ \text{for}\ \ \ \ r=r_{\gamma}\  ,
\end{equation}
or equivalently [see Eq. (\ref{Eq9})]
\begin{equation}\label{Eq14}
4\pi r^2p=1-{{3m(r)}\over{r}}\ \ \ \ \text{for}\ \ \ \ r=r_{\gamma}\  .
\end{equation}

Taking cognizance of the Einstein equations (\ref{Eq2}) and (\ref{Eq3}) and using 
the conservation equation
\begin{equation}\label{Eq15}
T^{\mu}_{r ;\mu}=0\  ,
\end{equation}
one obtains the gradient relation
\begin{equation}\label{Eq16}
{{d}\over{dr}}(r^2p)={{r}\over{2\mu}}\Big[{\cal N}(\rho+p)+2\mu(-\rho+p)\Big]\
\end{equation}
for the dimensionless pressure function $r^2p(r)$. 
From Eqs. (\ref{Eq9}), (\ref{Eq10}), (\ref{Eq13}), (\ref{Eq14}), and (\ref{Eq16}) one finds the gradient 
relation \cite{Hodrole}
\begin{equation}\label{Eq17}
\Big[{{d{\cal N}}\over{dr}}\Big]_{r=r_{\gamma}}={{2}\over{r_{\gamma}}}\big[1-8\pi r^2_{\gamma}(\rho+p)\big]\
\end{equation}
at the radial locations of the null circular geodesics. 

Our main goal is to determine the physical and mathematical properties of isotropic curved spacetimes that possess 
radial intervals $[r_-,r_+]$ with a continuum (an infinite number) of closed light rings. 
These unique radial intervals are characterized by the property (\ref{Eq13}) with the gradient relation 
\begin{equation}\label{Eq18}
{{d{\cal N}}\over{dr}}=0\ \ \ \ \text{for all}\ \ \ \ r_{\gamma}\in[r_-,r_+]\  ,
\end{equation}
or equivalently [see Eq. (\ref{Eq17})]
\begin{equation}\label{Eq19}
8\pi r^2(\rho+p)=1\ \ \ \ \text{for all}\ \ \ \ r_{\gamma}\in[r_-,r_+]\  .
\end{equation}
We have therefore proved that a radial interval $[r_-,r_+]$ that contains an {\it infinite} sequence (a continuum) of null 
circular geodesics is characterized by the functional relations (\ref{Eq10}), (\ref{Eq14}), and (\ref{Eq19}) 
for all $r_{\gamma}\in[r_-,r_+]$. 

It is interesting to note that one deduces from Eqs. (\ref{Eq5}), (\ref{Eq7}), 
(\ref{Eq11}), (\ref{Eq12}), and (\ref{Eq13}) that the special radial intervals $[r_-,r_+]$ that contain the infinite sequences (the continua) of null 
circular geodesics cannot extend all the way to the origin and to spatial infinity. 
In particular, light rings are characterized by the inequalities \cite{Notedec}
\begin{equation}\label{Eq20}
r_{\gamma}>0\ \ \ \ \text{and}\ \ \ \  r_{\gamma}<\infty\  ,
\end{equation}
which immediately imply the relations 
\begin{equation}\label{Eq21}
r_->0\ \ \ \ \text{and}\ \ \ \  r_+<\infty\
\end{equation}
for the boundaries of the special radial intervals that contain the infinite sequences of closed light rings in the 
curved spacetime (\ref{Eq1}).

One therefore concludes that our physically intriguing spacetimes possess a central core 
of finite radius $r_->0$ which supports a 
spherical shell $r_{\gamma}\in[r_-,r_+]$ of matter 
that contains an infinite number (a continuum) of light rings which 
are all characterized by the three functional relations (\ref{Eq10}), (\ref{Eq14}), and (\ref{Eq19}). 

\subsection{Functional expressions for the energy density and the pressure within light disks}
In the present subsection we shall explicitly determine the radially dependent functional expressions of 
the energy density $\rho(r)$ and the pressure $p(r)$ that characterize 
the self-gravitating isotropic matter fields within the special radial interval $[r_-,r_+]$ that contains the continuum 
of closed light rings. 

Differentiating both sides of Eq. (\ref{Eq14}) one obtains the relation [see Eq. (\ref{Eq10})] 
\begin{equation}\label{Eq22}
12\pi r^2p+4\pi r^3{{dp}\over{dr}}=1-12\pi r^2\rho\ \ \ \ \text{for all}\ \ \ \ r\in[r_-,r_+]\  ,
\end{equation}
which, taking cognizance of Eq. (\ref{Eq19}), yields the remarkably compact functional relation 
\begin{equation}\label{Eq23}
{{dp}\over{dr}}=-{{1}\over{8\pi r^3}}\ \ \ \ \text{for all}\ \ \ \ r\in[r_-,r_+]\  .
\end{equation}
From Eq. (\ref{Eq23}) one finds the radial functional behavior 
\begin{equation}\label{Eq24}
p(r)={{1}\over{16\pi r^2}}-\alpha\ \ \ \ \text{for all}\ \ \ \ r\in[r_-,r_+]\
\end{equation}
of the isotropic pressure function within the special radial interval $[r_-,r_+]$, where $\alpha$ is a constant. 
Substituting Eq. (\ref{Eq24}) into Eq. (\ref{Eq19}) one obtains the radially-dependent relation  
\begin{equation}\label{Eq25}
\rho(r)={{1}\over{16\pi r^2}}+\alpha\ \ \ \ \text{for all}\ \ \ \ r\in[r_-,r_+]\
\end{equation}
for the density function of the self-gravitating matter fields. 

We shall now show that the integration constant $\alpha$ in Eqs. (\ref{Eq24}) and (\ref{Eq25}) 
is uniquely determined by the radius $r_-$ and the gravitational mass \cite{Notemcore}
\begin{equation}\label{Eq26}
m_{\text{c}}\equiv m(r=r_-)\
\end{equation}
of the central core that supports the special shell $[r_-,r_+]$ with the infinite sequence (the continuum) 
of null circular geodesics. 
To this end, we shall first substitute the radially-dependent density expression (\ref{Eq25}) 
into the integral relation [see Eqs. (\ref{Eq10}) and (\ref{Eq26})] 
\begin{equation}\label{Eq27}
m(r)=m_{\text{c}}+\int_{r_-}^{r} 4\pi x^{2}\rho(x)dx\ \ \ \ \text{for}\ \ \ \ r\geq r_-\  ,
\end{equation}
which yields the relation 
\begin{equation}\label{Eq28}
m(r)=m_{\text{c}}+{1\over4}(r-r_-)+{{4\pi\alpha}\over{3}}(r^3-r^3_-)\ \ \ \ \text{for}\ \ \ \ r\in[r_-,r_+]\  .
\end{equation}

Substituting Eqs. (\ref{Eq24}) and (\ref{Eq28}) into Eq. (\ref{Eq14}) one finds the dimensionless mass-to-radius 
ratio 
\begin{equation}\label{Eq29}
{{m_{\text{c}}}\over{r_-}}={1\over4}+{{4\pi\alpha}\over{3}}r^2_-\
\end{equation}
of the central supporting core, or equivalently
\begin{equation}\label{Eq30}
\alpha={{3}\over{4\pi r^3_-}}\cdot\big(m_{\text{c}}-{1\over4}r_-\big)\  .
\end{equation}

\subsection{Energy conditions and the compactness of the central core}
In the present subsection we shall show that physically motivated requirements, 
like the strong energy condition and the dominant energy condition \cite{Bekreg}, 
yield explicit lower and upper bounds on the dimensionless compactness parameter 
\begin{equation}\label{Eq31}
{\cal C}_{\text{c}}\equiv{{m_{\text{c}}}\over{r_-}}\
\end{equation}
which characterizes the inner core that supports the special 
radial interval $[r_-,r_+]$ with the continuum of null circular geodesics. 

We first point out that, assuming that the self-gravitating matter fields 
respect the dominant energy condition \cite{Bekreg}, 
\begin{equation}\label{Eq32}
0\leq|p|\leq\rho\  ,
\end{equation}
one deduces from Eqs. (\ref{Eq24}) and (\ref{Eq25}) the relation 
\begin{equation}\label{Eq33}
\alpha\geq0\  ,
\end{equation}
which yields the lower bound [see Eqs. (\ref{Eq30}) and (\ref{Eq31})]
\begin{equation}\label{Eq34}
\text{Dominant energy condition}\ \ \ \ \Longrightarrow\ \ \ \ {\cal C}_{\text{c}}\geq{1\over4}\
\end{equation}
on the compactness parameter of the central supporting core. 

In addition, assuming that the matter fields respect the strong energy condition \cite{Bekreg}, 
\begin{equation}\label{Eq35}
\rho+3p\geq0\  ,
\end{equation}
one finds from Eqs. (\ref{Eq24}) and (\ref{Eq25}) the relation 
\begin{equation}\label{Eq36}
\alpha\leq{{1}\over{8\pi r^2}}\ \ \ \ \text{for all}\ \ \ \ r\in[r_-,r_+]\  ,
\end{equation}
which yields the upper bound [see Eqs. (\ref{Eq30}) and (\ref{Eq31})] \cite{Noterpm}
\begin{equation}\label{Eq37}
\text{Strong energy condition}\ \ \ \ \Longrightarrow\ \ \ \ {\cal C}_{\text{c}}\leq{{1}\over{4}}+
{{r^2_-}\over{6r^2_+}}\
\end{equation}
on the dimensionless compactness parameter that characterizes the central supporting core. 

\subsection{Isotropic light disks with a vanishing external pressure}
It is physically interesting to note that if one assumes that the outer edge of the special 
interval $[r_-,r_+]$, which contains the continuum of null circular geodesics, 
is also the outer edge of the entire compact self-gravitating matter configuration with the 
characteristic property \cite{Notentt}
\begin{equation}\label{Eq38}
p(r=r_+)=0\  ,
\end{equation}
then one finds from Eq. (\ref{Eq24}) the compact expression
\begin{equation}\label{Eq39}
r_+=\sqrt{{{1}\over{16\pi\alpha}}}\ \ \ \ \text{for}\ \ \ \ p(r=r_+)=0\  ,
\end{equation}
or equivalently [see Eqs. (\ref{Eq30}) and (\ref{Eq31})]
\begin{equation}\label{Eq40}
{\cal C}_{\text{c}}={1\over4}+{{r^2_-}\over{12r^2_+}}\ \ \ \ \text{for}\ \ \ \ p(r=r_+)=0\  .
\end{equation}
Note that the analytically derived expression (\ref{Eq40}) is consistent with the requirements (\ref{Eq34}) and 
(\ref{Eq37}) that follow from the dominant and the strong energy conditions. 

\section{Summary and discussion}
The Einstein-matter field equations of general relativity predict the existence closed 
light rings (null circular geodesics) in highly curved spacetimes of black holes and 
horizonless compact objects. 
Interestingly, recent observational studies \cite{Aki} support this physically important prediction.

Light rings in curved spacetimes are usually characterized by {\it discrete} radii \cite{Bar,Chan,Shap,Hodub,Herne,Hodrec,YY}. 
Motivated by the important roles that null circular geodesics play in the physics of 
non-trivial curved spacetimes (see \cite{Bar,Chan,Shap,Herne,Hodns,Hodrec,YY,Aki,Mash,Goeb,Hod1,Dec,Hodhair,Hodfast,
YP,Hodm,Hodub,Lu1,Hodlwd,Pod,Ame,Ste} and references therein), 
in the present paper we have raised the following physically intriguing question: 
Is it possible to build curved spacetimes that contain an infinite number (a {\it continuum}) of light rings? 

Using the non-linearly coupled Einstein-matter field equations we have explicitly proved 
that the answer to the above stated question is `yes'. 
The main analytical results derived in this paper and their physical implications are as follows:

(1) We have revealed, for the first time, that there are well behaved solutions of the Einstein-matter field equations that describe 
non-trivial isotropic curved spacetimes with an infinite number of null circular geodesics. 
These physically interesting spacetimes possess a central core of a finite radius $r_->0$ 
that supports a spherical shell $r_{\gamma}\in[r_-,r_+]$ which contains a continuum 
of light rings that are all characterized by the functional 
relations (\ref{Eq10}), (\ref{Eq14}), and (\ref{Eq19}). 

(2) We have proved that the self-gravitating isotropic matter fields in the special 
interval $[r_-,r_+]$, which contains the continuum of light rings, 
are characterized by the radially dependent functional relations 
[see Eqs. (\ref{Eq24}), (\ref{Eq25}), (\ref{Eq28}), (\ref{Eq30}), and (\ref{Eq31})]
\begin{equation}\label{Eq41}
p(r)=p_{\text{T}}(r)={{1}\over{16\pi r^2}}-{{3}\over{4\pi r^2_-}}\cdot\big({\cal C}_{\text{c}}-{1\over4}\big)
\ \ \ \ \text{for}\ \ \ \ r\in[r_-,r_+]\  ,
\end{equation}
\begin{equation}\label{Eq42}
\rho(r)={{1}\over{16\pi r^2}}+{{3}\over{4\pi r^2_-}}\cdot\big({\cal C}_{\text{c}}-{1\over4}\big)\ \ \ \ \text{for}\ \ \ \ r\in[r_-,r_+]\  ,
\end{equation}
and
\begin{equation}\label{Eq43}
m(r)={1\over4}r+{{r^3}\over{r^2_-}}\cdot\big({\cal C}_{\text{c}}-{1\over4}\big)\ \ \ \ \text{for}\ \ \ \ r\in[r_-,r_+]\  ,
\end{equation}
where $r_-$ and ${\cal C}_{\text{c}}=m_{\text{c}}/r_-$ are respectively the radius of the central supporting core and its dimensionless compactness parameter. 

(3) From the analytically derived expression (\ref{Eq43}) one finds that the radially dependent compactness function
\begin{equation}\label{Eq44}
{\cal C}(r)\equiv{{m(r)}\over{r}}\  ,
\end{equation}
which characterizes the self-gravitating isotropic matter configurations in the interval $[r_-,r_+]$, is 
given by the dimensionless functional relation 
\begin{equation}\label{Eq45}
{\cal C}(r)={1\over4}+{{r^2}\over{r^2_-}}\cdot\big({\cal C}_{\text{c}}-{1\over4}\big)
\ \ \ \ \text{for}\ \ \ \ r\in[r_-,r_+]\  .
\end{equation}

From Eq. (\ref{Eq45}) one deduces that the no-horizon condition $\mu(r)>0$ 
[or equivalently ${\cal C}(r)<1/2$, see Eqs. (\ref{Eq9}) and (\ref{Eq44})] yields the inequality
\begin{equation}\label{Eq46}
{{r}\over{r_-}}<\sqrt{{{1}\over{4{\cal C}_{\text{c}}-1}}}\ \ \ \ \text{for all}\ \ \ \ r\in[r_-,r_+]\  ,
\end{equation}
which implies the dimensionless upper bound \cite{Noterpm2}
\begin{equation}\label{Eq47}
{{r_+}\over{r_-}}<\sqrt{{{1}\over{4{\cal C}_{\text{c}}-1}}}\
\end{equation}
on the outer radius of the special interval that contains the continuum of light rings. 

(4) We have proved that, for self-gravitating isotropic matter configurations that respect the 
dominant energy condition and the strong energy condition \cite{Bekreg}, the dimensionless 
compactness parameter of the central supporting core 
is bounded by the two inequalities 
[see Eqs. (\ref{Eq31}), (\ref{Eq34}), and (\ref{Eq37})] \cite{Notelu,Noterr,Notesg}
\begin{equation}\label{Eq48}
{{1}\over{4}}\leq{\cal C}_{\text{c}}\leq{{5}\over{12}}\  .
\end{equation}



\bigskip
\noindent {\bf ACKNOWLEDGMENTS}

This research is supported by the Carmel Science Foundation. I thank
Yael Oren, Arbel M. Ongo, Ayelet B. Lata, and Alona B. Tea for
stimulating discussions.


\begin{thebibliography}{99}
\bibitem{Noteunit} We shall use natural units in which $G=c=1$.
\bibitem{NoteMM} Here $M$ is the total (asymptotically measured) mass of the spacetime.
\bibitem{Notedec} Here we have used the dominant energy condition [see Eq. (\ref{Eq32}) below].
\bibitem{Notemcore} Note that $m_{\text{c}}$ as defined by Eq. (\ref{Eq26}) is the mass of the central core that 
supports the radial interval $[r_-,r_+]$ with the infinite sequence of light rings.
\bibitem{Noterpm} Here we have used the fact that, in the interval $r\in[r_-,r_+]$, 
the radial expression ${{1}/{8\pi r^2}}$ is minimized for $r=r_+$.
\bibitem{Notentt} Note that the relation (\ref{Eq38}) is not a necessary requirement. 
In particular, if $p(r=r_+)\neq0$ then the unique interval $[r_-,r_+]$, which contains the continuum of light rings, 
is surrounded by a larger spherical region that contains non-vacuum matter fields with the characteristic 
property $p(r\to\infty)\to0$ [see Eqs. (\ref{Eq12}) and (\ref{Eq32})].
\bibitem{Noterpm2} Here we have used the fact that, in the interval $r\in[r_-,r_+]$, 
the dimensionless ratio $r/r_-$ is maximized for $r=r_+$.
\bibitem{Notelu} Note that the lower/upper bounds in (\ref{Eq48}) follow respectively 
from the dominant/strong energy conditions [see Eqs. (\ref{Eq34}) and (\ref{Eq37})].
\bibitem{Noterr} Here we have used the dimensionless inequality $r_-/r_+\leq1$ [see Eq. (\ref{Eq37})].
\bibitem{Notesg} Note that the upper bound ${\cal C}_{\text{c}}\leq{{1}/{4}}+{{r^2_-}/{6r^2_+}}$, 
which follows from the strong energy condition (\ref{Eq35}), is stronger than the bound 
${\cal C}_{\text{c}}\leq{{1}/{4}}+{{r^2_-}/{4r^2_+}}$ which follows from the no-horizon condition (\ref{Eq47}).
\end{thebibliography}
\end{document}